**High-Low Refractive Index Stacks for Broadband Antireflection Coatings for Multijunction Solar Cells**

*GuoJiao Hou, Iván García, Ignacio Rey-Stolle\**

Instituto de Energía Solar – Universidad Politécnica de Madrid
ETSI de Telecomunicación, Avda. Complutense 30, 28040 Madrid, SPAIN
E-mail: ignacio.reystolle@upm.es (I. Rey-Stolle)

**Abstract**: Antireflection coatings are an interesting challenge for multijunction solar cells due to their broadband spectrum absorption and the requirement of current matching of each subcell. A new design for multijunction solar cell antireflection coatings is presented in this work in which alternative high and low index materials are used to minimize the reflection in a broadband (300nm-1800nm). We compared the short circuit current density of high-low refractive index stacks designs with optimum double-layer antireflection coatings by considering two optical materials combinations ($MgF_2$/ZnS and $Al_2O_3$/$TiO_2$) for the AM0 and AM1.5D spectra. The calculations demonstrate that for lattice-matched triple-junction solar cells and inverted metamorphic quadruple-junction solar cells, high-low refractive index stacks outperform the optimum double-layer antireflection coatings. The new design philosophy requires no new optical materials because only two materials are used and exhibits excellent performance in broadband spectra. The angle performance of these antireflection coatings is slightly better than classical double-layers whereas the analysis for thickness sensitivity shows that small deviations from deposition targets only slightly impact the performance of antireflection coatings. Finally, some technical solutions for depositing these high-low refractive index multilayers are discussed.

# 1. Introduction

Multijunction solar cells have the potential to realize higher efficiencies than the Shockley-Queisser limit for single junction cells due to the absorption of a broad range of the solar spectrum, whilst minimizing thermalisation losses (Shockley and Queisser, 1961). Multijunction solar cells are particularly suitable for space applications because of their high reliability, high power-to-mass ratio (W/m$^2$·kg) and excellent radiation hardness (Zhang et al., 2017; Kawakita et al., 2016). In addition, the development of advanced concentrator systems shows promise for attractive reductions in the cost of utility-scale terrestrial solar power plants.

Antireflection coatings (ARC) are vital components for multijunction solar cells, where the situation is demanding since solar power is spread in a broadband spectral range and, typically, reaches the solar cell in a wide distribution of angles of incidence. The ideal ARCs should exhibit minimal reflection in a broad spectral range (300-1800nm) and even performance among the different subcells because any imbalance may increase the current mismatch. Furthermore, the materials used in ARCs should be accessible, reliable, durable and suited to be deposited uniformly over large areas.

The classic approach to suppress the reflection in a solar cell is a step-down structure in which the refractive index decreases in steps from that of the solar cell material to that of the outer medium or ambient, with the antireflective bandwidth increasing with the numbers of steps. For example, in state-of-the-art GaInP/Ga(In)As/Ge triple-junction solar cells a double-layer ARC (DLAR) formed by two materials (i.e. two steps) exhibits a low integrated reflection (<5%) from 350-1200nm (Aiken, 2000). It is true that the spectral response of the Ge subcell extends further down to 1850nm, but as this cell produces excess current, a poorer performance of the ARC in this range does not translate into a reduced short circuit current from triple-junction solar cell (Aiken, 2000). However, the situation in more advanced designs with 4, 5 or even 6 junctions is more critical and the ARC should perform equally well in each subcell band (Aiken, 2000). So an obvious alternative that has been explored to improve the ARC performance has been to increase the number of steps exploring new optical materials (Zhang et al., 2019; Sikder and Zaman, 2016; France et. al., 2014; Stetter et. al., 1976).

Nevertheless, such improvement of classic step-down ARCs sometimes struggles with the lack of materials with the required refractive indices. To sort this out Herpin equivalents were proposed (Epstein, 1952). In brief, the layer of unavailable refractive index can be substituted by a train of symmetrical layers according to Herpin equivalent equations (Epstein, 1952). However, the application of such Herpin layers will produce equivalent performance only at a single wavelength and is restricted by symmetrical combinations (three, five, or some odd number of layers) (Epstein, 1952). As a result of this limitation, Herpin equivalents have found few applications in ARCs for solar cells (Aiken, 2000; France et. al., 2014). Beyond thin-film approaches, novel ARCs based on plasmonics and nanostructures have been reported (Kim et al., 2019; Xi et al., 2007; Baryshnikova et al., 2016; Zhang et al., 2020; Abdelraouf et al., 2018; Huang et al., 2016; Prieto et al., 2009). However, these interesting ideas have been found limited application in the PV industry as they require very sophisticated (and expensive) manufacturing processes that are generally restricted to small areas.

The goal of this study is to stick to thin film-based ARC designs for their manufacturability and low cost as compared to other approaches based on nanostructures. In this respect, we aim at creating better ARC designs combining thin layers of only two optical materials, namely, a high-refractive index material with n~2.5 and a low refractive index material with n~1.5. This idea roots from the seminal paper by Southwell (Southwell, W.H., 1985), where a novel broadband ARC design for glass using very thin alternating layers with high- and low-index was reported (Southwell, 1985). Southwell's paper proved that an optical layer in a coating is equivalent to a combination of thin high and low index layers, with such equivalence being calculated by comparing their characteristic matrices. Southwell reported examples of different antireflection coatings (eg. a quintic index profile and a quarterwave ARC) on glass that could be successfully be replaced by high-low equivalent pairs. Even more so, he demonstrated that any arbitrary optical interference coating consisting of gradient-index layers can be substituted by a sequence of high- and low-index layers with equivalent performance (Southwell, 1985). Therefore, the high- and low-index layers can be used to generate high performance ARCs in broadband spectrum and do not need to consider the limitation of materials with certain refractive index as in gradient-index ARCs.

In this paper, we apply Southwell's strategy to produce a new algorithm to calculate optimum ARCs based on High-Low Refractive Index Stacks (HLIS) for GaInP/Ga(In)As/Ge triple-junction solar cells and inverted metamorphic quadruple-junction GaInP/GaAs/GaInAs/GaInAs solar cells for different incident media (air and glass) and spectra (AM0 and AM1.5D). We conduct these calculations for two pairs of commonly used optical materials –$MgF_2$/ZnS and $Al_2O_3$/$TiO_2$ – and compare their performance with the best conventional DLAR. Finally, we contextualize the photocurrent gains obtained with HLIS designs and the complexity and practicality of the coatings devised.

## 2. Simulation Approach

The simulations performed in this study have been performed on two representative solar cell designs, namely, a state-of-the-art lattice-matched triple-junction $Ga_{0.50}In_{0.50}P$/$Ga_{0.99}In_{0.01}As$/Ge solar cell (3JLM) (King et al., 2009; Bett et al., 2009; Barrutia et al., 2018), and an inverted metamorphic four-junction $Ga_{0.5}In_{0.5}P$/GaAs /$Ga_{0.72}In_{0.28}As$/$Ga_{0.43}In_{0.57}As$ device (4JIMM) (France et al., 2014, 2016; King et al., 2012). Obviously, the internal quantum efficiency (IQE) of each design determines the maximum achievable photocurrent so the performance of any ARC design can be benchmarked against such ideal $J_{SC}$ obtained by convoluting the IQE with the reference spectrum. Therefore, we can define the *Achievable Photocurrent Fraction* (*APF*) as a key figure of merit (**Equation 1**):

$$APF = \frac{J_{SC}}{J_{SC,IQE}} \qquad (1)$$

with $J_{SC}$ being the short circuit current of the multijunction solar cell with series connected subcells achieved with certain ARC and $J_{SC,IQE}$ the ideal photocurrent achievable without any reflection losses. The IQE of both designs is depicted in *Fig 1*, where two versions have been optimized for each design since we will perform calculations for two spectra, AM0 and AM1.5D. As shown by *Fig 1(a)-(d)*, the IQEs of the subcells are well balanced in all cases so *APF* is an excellent metric on how a given ARC design approaches ideal performance. The $J_{SC,IQE}$ of each subcell is shown in the figure as well.

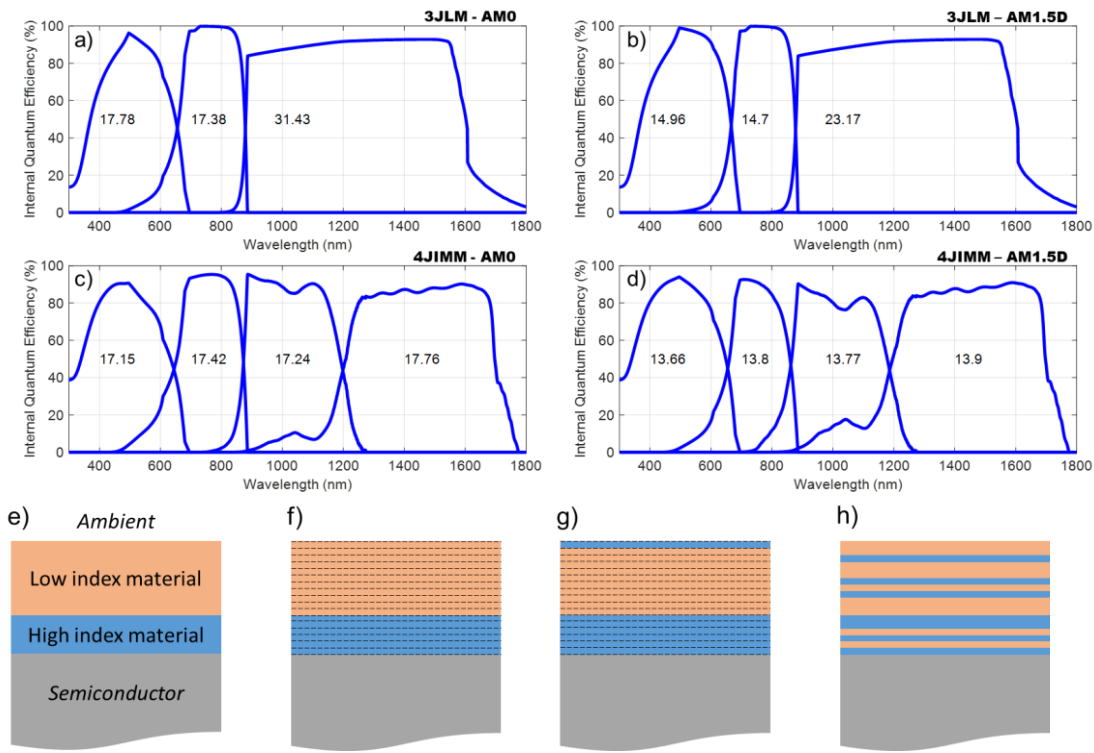

**Fig 1**. Internal quantum efficiency of the solar cell designs used in this study. (a) 3JLM for AM0; (b) 3JLM for AM1.5D; (c) 4JIMM for AM0; (d) 4JIMM for AM1.5D. Schematic of the optimization algorithm to calculate the optimum HLIS (e) Starting DLAR and optical boundary conditions; (f) slicing of the DLAR into thin layers; (g) first iteration, i.e. topmost thin slice flipped; (h) resulting HLIS design after the last pass.

In *Fig 1 (e)-(h)*, the schematic of HLIS optimization algorithm is shown. The starting point is a DLAR design based on a high refractive index material (shown in blue) and a low refractive index material (shown in orange). The DLAR design is calculated for two spectra AM0 and AM1.5D and the detail of calculation can be found in *Fig 2*. Then we slice that into thin layers with equal thickness. We change the state of a thin layer one at a time (shown in Fig1(g) from low index to high index) and calculate the subsequent *APF*. If the *APF* improves, the change remains, otherwise we restore the ARC to its prior state. After few passes, we get the final result of the HLIS design. More detail about this algorithm is described below. The simulation environment devised to calculate the $J_{SC}$ of a multijunction solar cell is quite straightforward. For a certain ambient and ARC design, we calculate the spectral reflectance at the top surface and the optical absorption in each subcell using the Scattering Matrix Method as formulated by Centurioni (Centurioni, 2005) and then the External quantum efficiency (EQE) of the structure is calculated using the Hovel model (Rey-Stolle et al., 2016) to eventually obtain the $J_{SC}$ in each subcell by convoluting the corresponding EQE and the spectrum under consideration. The multijunction $J_{SC}$ is the minimum $J_{SC}$ among the subcells. Using this simulation backbone, we have implemented an algorithm inspired in the reference (Southwell, 1985) that, starting from a classic DLAR, finds a coating with the same thickness, using the same materials, but with an improved antireflection performance measured as a higher value of its *APF*. The process is as follows:

    (1) Calculate the optimum DLAR for the material pair

    (2) Evaluate the figure of merit (*APF*) of the optimum DLAR

(3) The resulting DLAR is sliced into a stack of thin layers of equal thickness; in our study we have considered thin layers of 5nm and 2.5nm. Conceptually, thinner layers could be used, but we have limited our choice to those two values bearing in mind the practical manufacturability of the coatings with common ARC deposition tools.

(4) Change the material of each layer (from high to low or *vice versa*) one at a time and reevaluate the *APF*. If the performance is better in the flipped state, retain the change; otherwise restore it.

(5) If, after testing all the thin layers (a single pass), the merit function has improved, go to step (4) for another pass; otherwise end.

In this algorithm, what we get is the optimum result of a fixed film thickness and a refinement can be achieved by using a thinner slice thickness.

For the calculation of the optimum DLAR –step (1)– a classic brute force approach has been followed where the thickness of the two materials involved has been varied over reasonable ranges and the performance of the resulting ARC (i.e. the $J_{SC}$ of the multijunction) has been calculated. The calculation of $J_{SC}$ is shown in **Equation (2)**, where $q$ is electron charge, $[\lambda_1, \lambda_2]$ is the wavelength range (300nm-1800nm); $F(\lambda)$ is the spectral photon flux density and $R(\lambda)$ is the reflectance. A different reflectance is achieved with the changes in the thicknesses of ARC layers and $J_{SC}$ evolves correspondingly as shown in *Fig 2*.

$$J_{SC} = q \int_{\lambda_1}^{\lambda_2} F(\lambda)[1 - R(\lambda)]IQE(\lambda)d\lambda \qquad (2)$$

This allows to draw false color maps where photocurrent maxima can be identified. Two examples of such false color contour plots are included in *Fig 2*. The *Fig 2(a)* is the photocurrent for a triple-junction solar cell calculated for AM1.5D, ambient glass and when $MgF_2$/ZnS materials are used in the DLAR. This figure reflects an important feature for our study as is the presence of an absolute $J_{SC}$ maximum at $MgF_2$=0 nm and ZnS=65 nm with 14.43 mA/cm$^2$ (note that for this particular case the calculations are done for ambient glass), and a secondary local maximum at $MgF_2$=240nm and ZnS=60 nm with similar performance ($J_{SC}$ = 14.35 mA/cm$^2$). This secondary local maximum provides an extra starting point for our algorithm that, in some cases, converges to a better final solution as will be shown in the next section. Therefore, step (1) in the algorithm involves the calculation of the optimum DLAR as well as the identification of possible local maxima of similar performance. Analogously, *Fig 2(b)* is the photocurrent of a triple-junction solar cell calculated for the case of AM1.5D spectrum, ambient glass, and using $Al_2O_3/TiO_2$ materials in the DLAR. The absolute $J_{SC}$ maximum is achieved at $Al_2O_3$=90nm and $TiO_2$=40 nm with 14.34 mA/cm$^2$. As shown in the figure, there is no secondary local maximum with similar performance in this case.

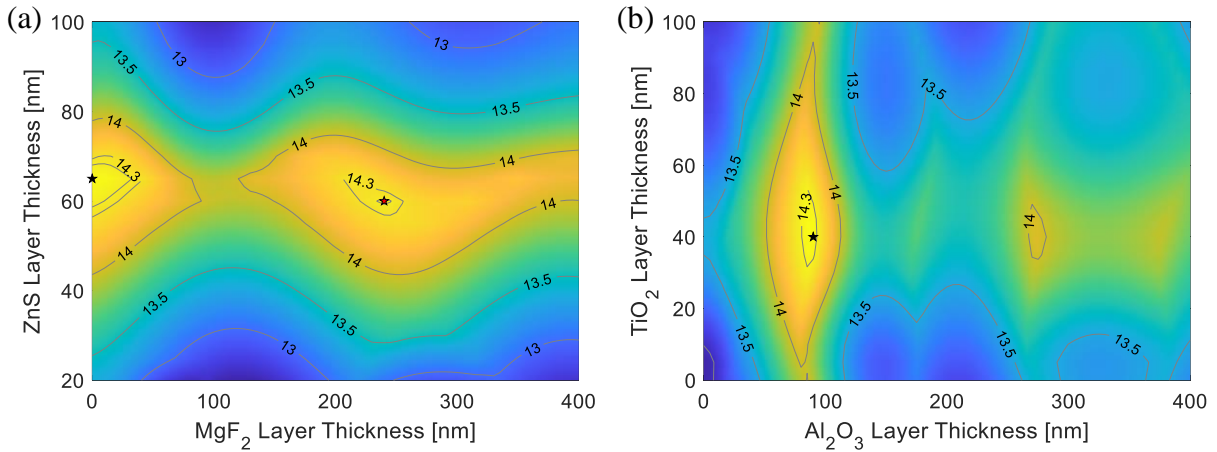

**Fig 2**. False color map of the photocurrent of the 3JLM vs. the thickness of the high and low index layers in the DLAR. (a)Calculated for ambient glass, spectrum AM1.5D and $MgF_2$ (low index), ZnS (high index). The black star symbol represents the absolute maximum ($J_{SC}$=14.43 mA/cm$^2$), whereas the red star symbol represents a secondary maximum with similar performance ($J_{SC}$=14.35 mA/cm$^2$) (b) Calculated for ambient glass, spectrum AM1.5D and $Al_2O_3$ (low index), $TiO_2$ (high index). The black star symbol represents the absolute maximum ($J_{SC}$=14.34 mA/cm$^2$) and no secondary local maximum is found.

In order to reflect the contexts where III-V multijunction solar cells are used today we have run the algorithm above described for designs optimized for AM0 (space applications) and AM1.5D normalized to 1000 W/m$^2$ (terrestrial high concentrator applications). Please note here that we have upscaled AM1.5D spectrum from the nominal 900 W/m$^2$ to 1000 W/m$^2$, following the usual convention in CPV intended to establish easier $J_{SC}$ comparisons with AM1.5G results. We have also considered two possible variants for the outer media, namely, air/vacuum –i.e. the case for space cells without a cover glass or concentrator cells without a secondary optical element– and glass –i.e. the case for space cells with a cover glass or concentrator cells embedded in a refractive secondary optical element. Moreover, we conduct these calculations for two pairs of commonly used optical materials, MgF$_2$/ZnS (frequently used in lab cells) and Al$_2$O$_3$/TiO$_2$ (mostly used in industry). Wavelength dependent optical constants have been taken from the Sopra database (Sopra) in the case of dielectrics, glass, air and vacuum. The refractive indices and extinction coefficients of semiconductors have been taken from specific reports in the literature (Adachi et al., 2013; Ochoa-Martínez et al., 2018). By sweeping this multiplicity of environments and variables we intend to determine whether high-low index stacks can provide an anti-reflecting solution of general applicability.

## 3. Results and Discussion

### 3.1. GaInP/Ga(In)As/Ge Triple-Junction Solar Cells

*Fig 3* shows the results of the simulation for spectra AM1.5D and both ambients, air and glass. As mentioned above, this would be a typical case for a concentrator multijunction solar cell. This figure contains several graphs arranged in two columns. The column on the left represents the comparison of the performance of the optimum DLAR vs. a High-Low index stack formed by thin steps of 5nm; whereas the column on the right does the same for thin layers of 2.5nm. In each graph the configuration of the ARC is represented by plotting the mean refractive index of the layer vs depth in the coating, with zero being the interface with the semiconductor and the final thickness the interface with the ambient and the total thickness of the coating. In such graphs, the thick red line corresponds to the DLAR (starting point for the optimization algorithm) whilst the thin blue line depicts the best High-Low stack reached for the given step thickness. The $J_{SC}$ and *APF* of each ARC are indicated next to the graph following the same color code, and the gains in $J_{SC}$ using the High-Low index stack are indicated in black ($\Delta J_{SC}$).

Panel a) in *Fig 3*, represents the case of using $MgF_2$ as low index material (n~1.4) and ZnS for high index material (n~2.4) spectrum AM1.5D and ambient glass. As shown in *Fig 2(a)*, the optimum DLAR in this case is formed by 0 nm of $MgF_2$ and 65 nm of ZnS; with a secondary local maximum of similar performance at 240 nm of $MgF_2$ and 60 nm of ZnS. Focusing on the main maxima, we must acknowledge that the DLAR shows a great performance with an *APF* = 98.2%, i.e. we only lose 1.8% of the maximum current reachable according to the IQE. However, the application of our algorithm produces a HLIS with a slight increase in $J_{SC}$ ($\Delta J_{SC}$= +0.47%) and thus an improvement of *APF* = 98.6%. Using the thinnest slices of 2.5 nm a minute further improvement is obtained (*APF* = 98.7%). Moving to the secondary maximum, the initial $J_{SC}$ is a bit worse but the improvement with the HLIS is notable ($\Delta J_{SC}$> +1.5%), reaching an *APF*=99.3% and thus a better photocurrent than in the case of the main maximum. Obviously, the DLAR design of the secondary maximum is thicker and thus gives more margin for the algorithm to squeeze the last bit of performance out of this combination of materials.

Panel b) in *Fig 3*, represents the case of using $Al_2O_3$ as low index material (n~1.4) and $TiO_2$ for high index material (n~2.1) for the same boundary conditions. These materials are a common choice in industry cells as a result of their durability and reliability, though they provide a slightly lower performance than $MgF_2$/ZnS. As shown in *Fig 2(b)*, the optimum DLAR in this case is formed by 90 nm of $Al_2O_3$ and 40 nm of $TiO_2$, with no secondary maximum providing similar performance. As noted, this DLAR performs slightly worse than the DLAR with $MgF_2$/ZnS with a $J_{SC}$=14.34 mA/cm$^2$ and an *APF*= 97.6%. However, the optimum HLIS found produces a remarkable boost in performance ($\Delta J_{SC}$> +1.8%), reaching the photocurrent achieved by the best designs based on $MgF_2$/ZnS. So, in this particular case, the use of a HLIS configuration in the ARC allows to reach lab performance with industry validated materials.

The lower section of *Fig 3*, represents analogous comparisons for ambient air. In this case we can observe the same trends but with lower intensity because no secondary local maximum with better performance can be found and the relatively thin thickness of the absolute maximum ARC design limits the optimization. HLIS designs outperform DLAR in all the situations and in the HLIS design with steps of 2.5nm provides little improvement over those of 5nm

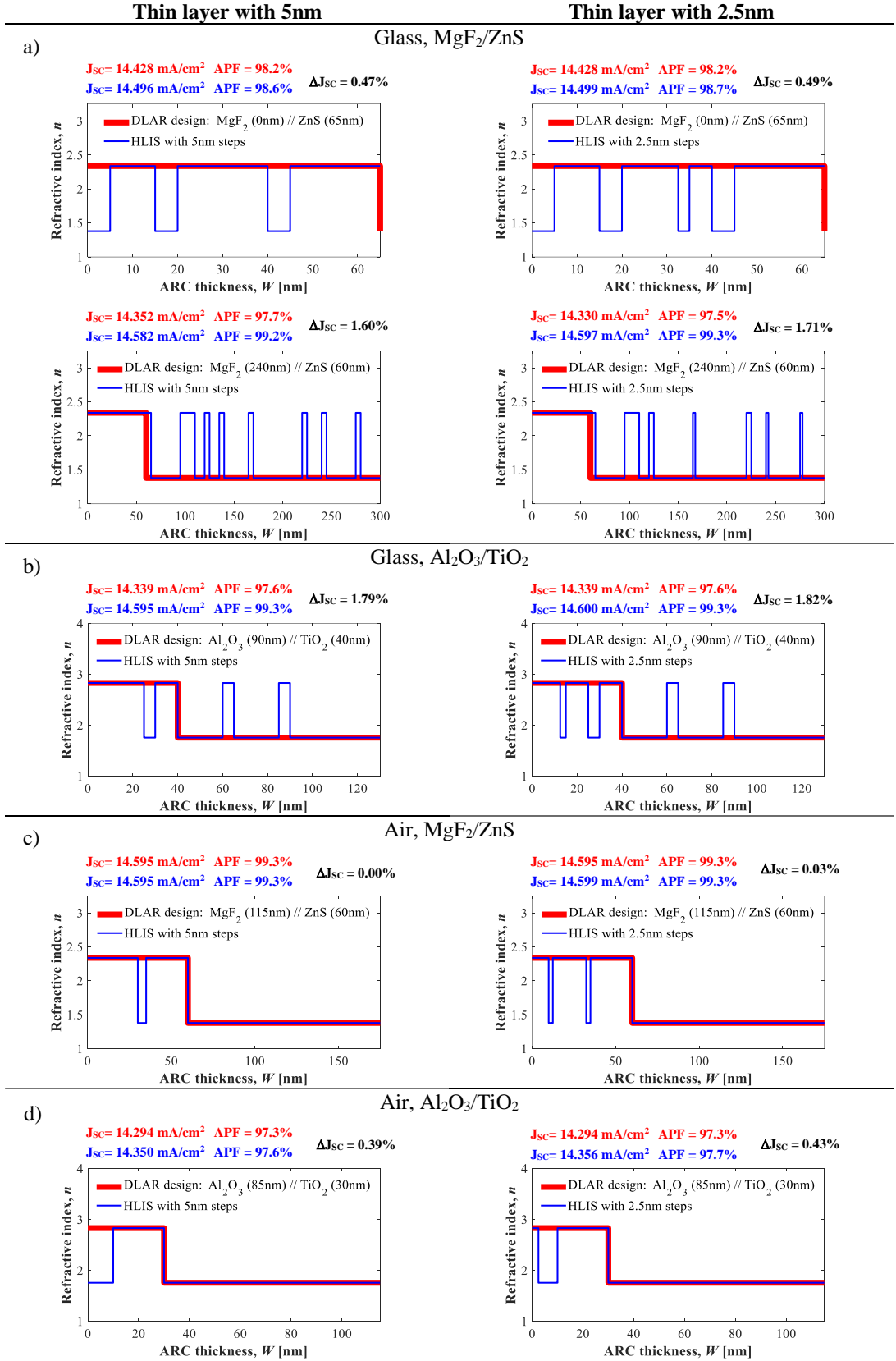

**Fig 3.** Comparison of High-Low refractive index stacks and Double-layer ARC designs in triple junction solar cells for spectrum AM1.5D. The left column presents the results for High-Low refractive index stacks with slices of 5nm whereas the right column is for 2.5nm. The ambient and materials are as indicated in each panel.

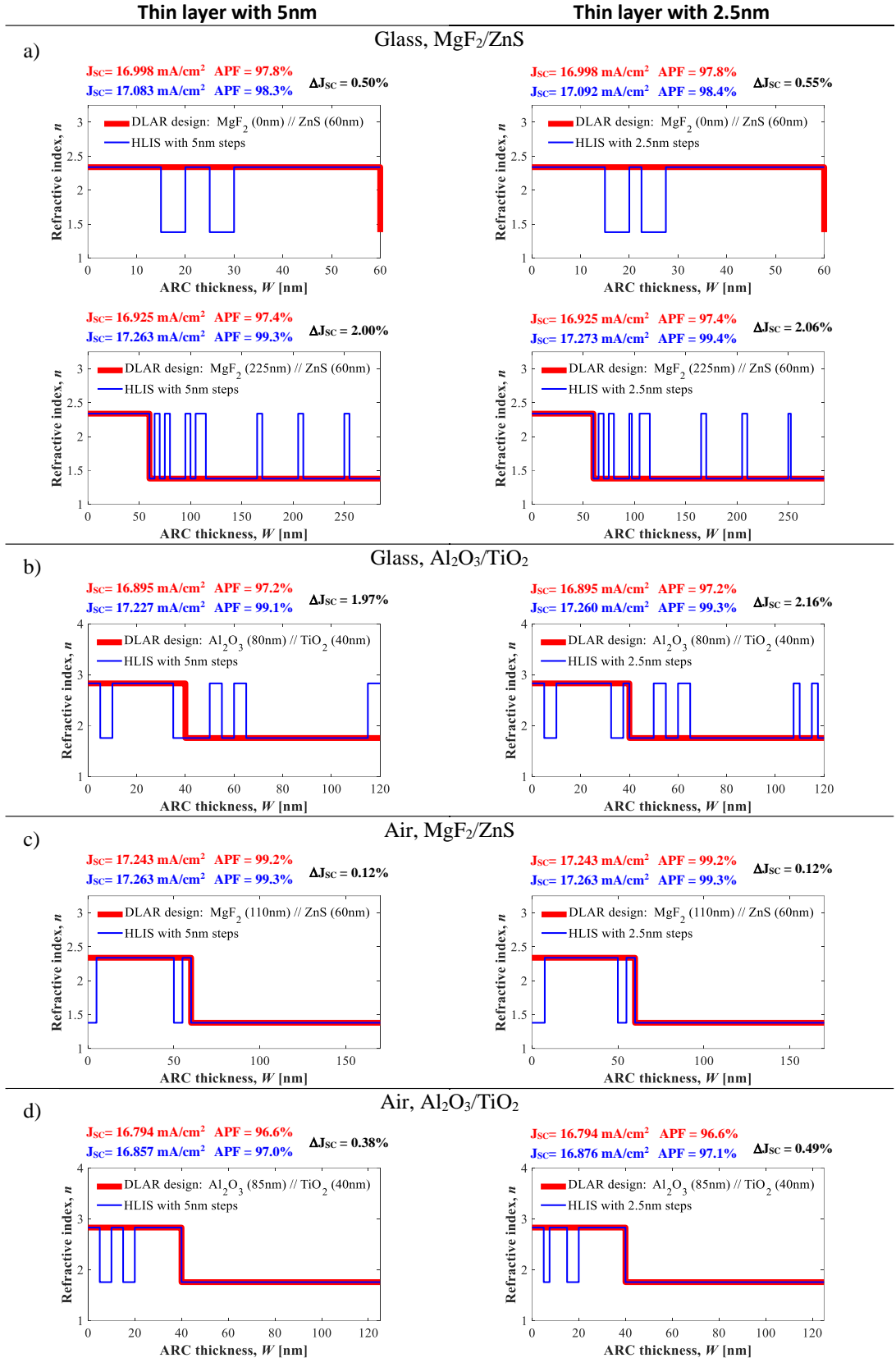

**Fig 4**. Comparison of High-Low refractive index stacks and Double-layer ARC designs in triple junction solar cells for spectrum AM0. The left column presents the results for High-Low refractive index stacks with slices of 5nm whereas the right column is for 2.5nm. The ambient and materials are as indicated in each panel.

*Fig 4* represents the same set of simulations for spectrum AM0, i.e. for space applications. Virtually the same behavior as in *Fig 3* can be observed. DLARs show a good performance – with high $J_{SC}$ values– which is improved by HLIS designs notably when the outer medium is glass and only marginally when the outer medium is air or vacuum.

**3.2. 4J Inverted Metamorphic (IMM) Solar Cells**

In the case of GaInP/Ga(In)As/Ge triple-junction solar cells discussed in the previous section, the need for a truly broadband ARC is partly mitigated by the excess current produced by the Ge bottom cell. The critical requirements for the ARC are those linked to the top and middle junctions (i.e. 350-900 nm) since the extra photocurrent available in the bottom cell can compensate for a modest performance of the ARC in the IR range. This is not the case in 4J IMM GaInP/GaAs/GaInAs/GaInAs solar cells where, as shown by the IQE in *Fig 1*, the spectral range covered by this structure is the same as 3JLM but with an even distribution of currents amid the four subcells. In this device the ARC must closely and evenly provide minimum reflectance from 350nm to ~1800nm.

*Fig 5* and *Fig 6* are analogous to *Fig 3* and *Fig 4* for 4J IMM GaInP/GaAs/GaInAs/GaInAs solar cells. Again, *Fig 5*, represents the results for spectrum AM1.5D, for both ambients (glass and air) and material pairs considered. A first general result is that the performance of the DLAR in all cases is lower than in the 3JLM case with *APF*~90-92%. In other words, when the spectral performance of the ARC is more demanding photocurrent loses get bigger (8-10%). However, when HLIS designs come into to play the gains are remarkable in all cases with $\Delta J_{SC}$~ as high as 4.6%. Moreover, most of the gain is obtained with steps of 5nm, with only minute increments when downsizing to steps of 2.5nm. Exactly the same trend can be deduced for *Fig 6* where the results for spectrum AM0 are gathered.

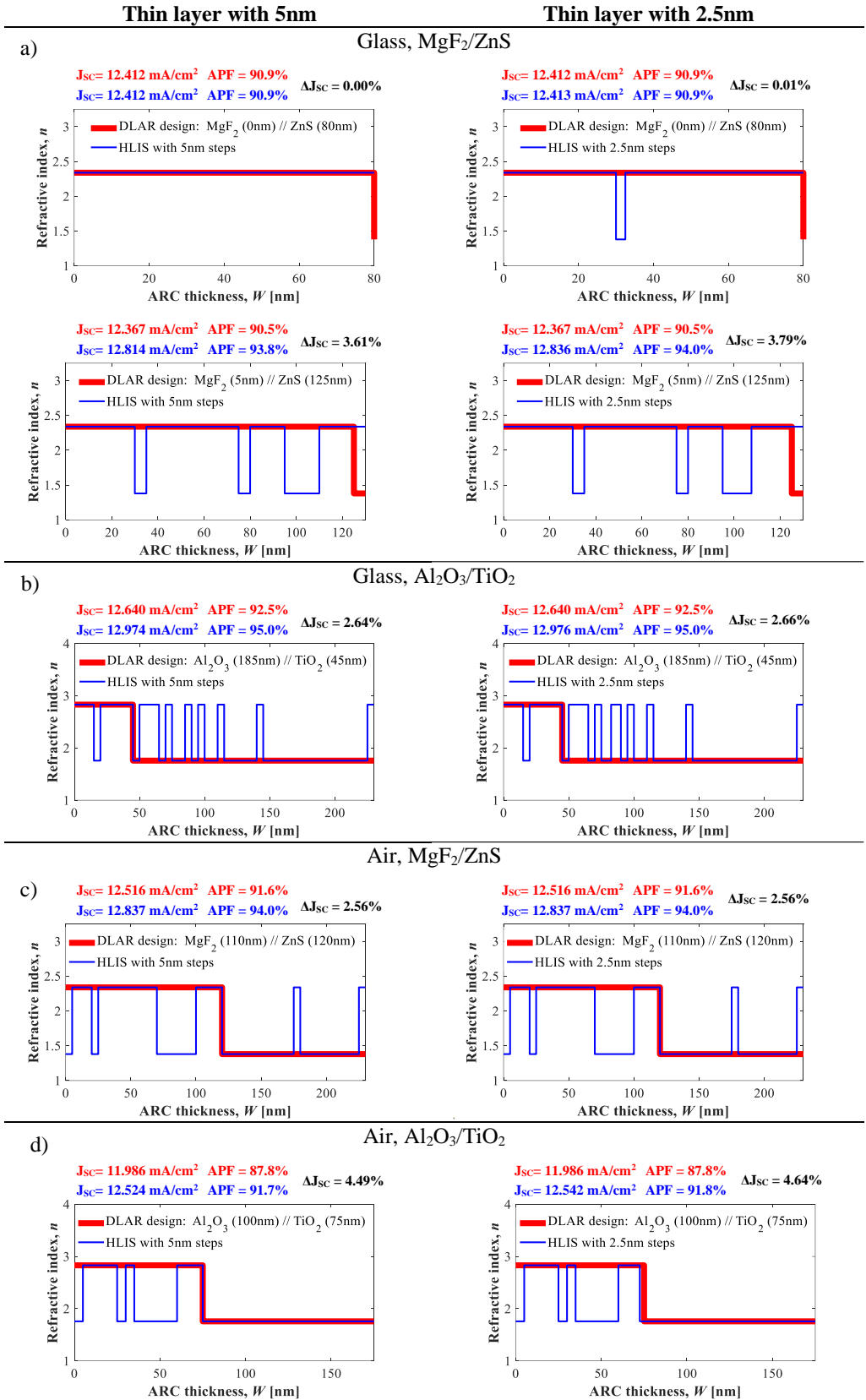

**Fig 5.** Comparison of High-Low refractive index stacks and Double-layer ARC designs in IMM four-junction solar cells for spectrum AM1.5D. The left column presents the results for High-Low refractive index stacks with slices of 5nm whereas the right column is for 2.5nm. The ambient and materials are as indicated in each panel.

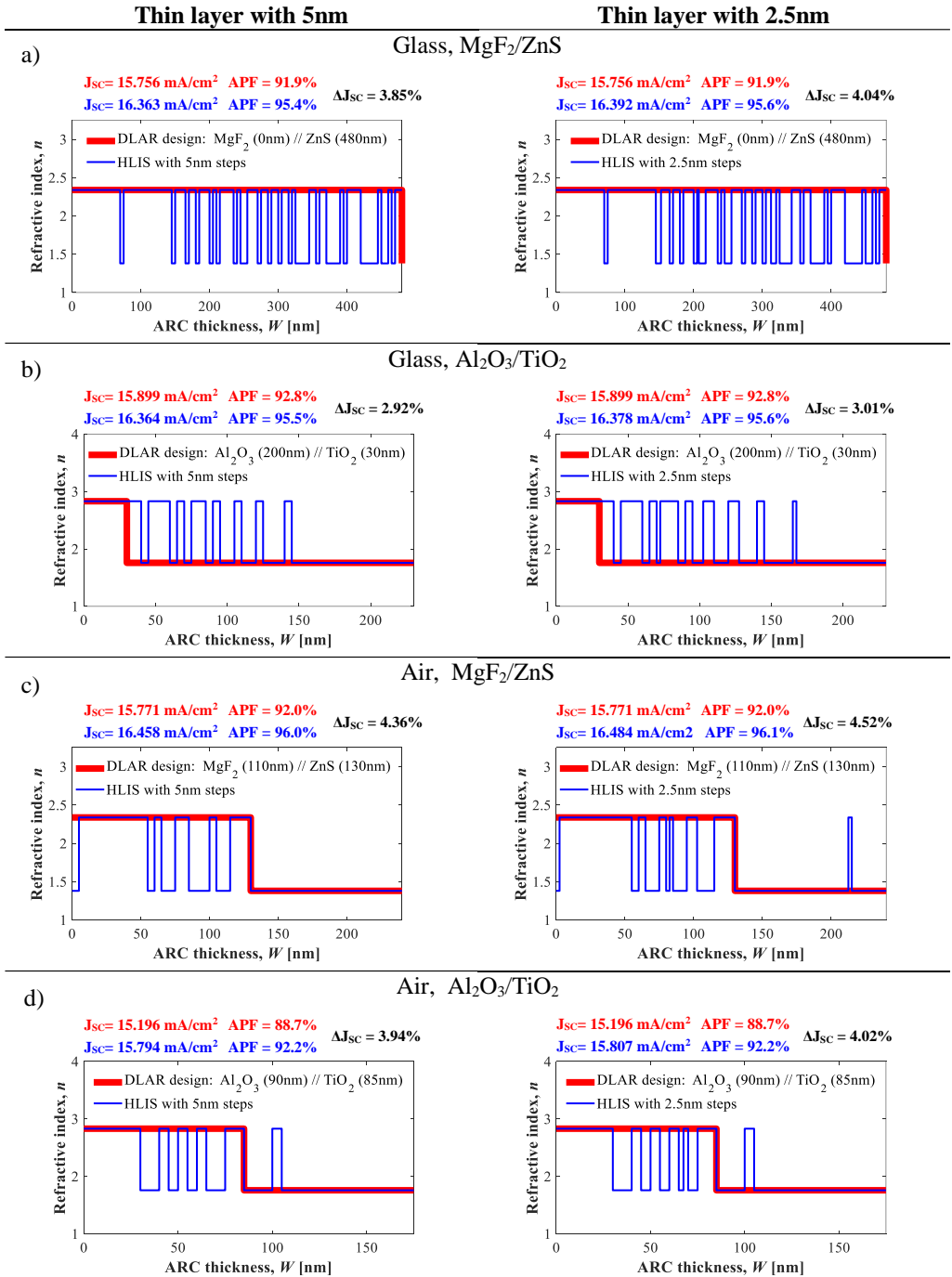

Fig 6. Comparison of High-Low refractive index stacks and Double-layer ARC designs in IMM four-junction solar cells for spectrum AM0. The left column presents the results for High-Low refractive index stacks with slices of 5nm whereas the right column is for 2.5nm. The ambient and materials are as indicated in each panel.

### 3.3. Discussion

According to calculations, we found the HLIS outperform the optimum DLAR for AM0 and AM1.5D spectra and compared to triple junction solar cells, quadruple junction solar cells achieved greater short circuit current gains because the current from each subcell is equivalent and the optimization can be implemented in the broadband spectrum. In the simulations, we used the commonly used materials in the lab cells and industry and the HLIS design exhibit excellent performance. Therefore, the HLIS design has application prospects. Of course, we should consider thin film deposition technology due to the thin layers in the design. In the next section, we will discuss issues that need to be considered in practice.

## 4. On the Practical Implementation of HLIS Designs

### 4.1. Impact of Deposition Tolerances on ARC Performance

In the previous section we have shown that HLIS outperform DLAR ARCs using the same materials and the same total thickness. Obviously, this comes at the expense of a higher complexity in the coating with (in extreme cases) tens of flips between high and low index materials. In experimental setups, oftentimes we cannot deposit the expected thickness due to inevitable experimental uncertainties and/or calibration errors. For example, Dong et al. (2018) reported that when depositing $TiO_2$, the deposition error was ~1% for a target thickness of 120 nm, whereas it soared to ~20% for a target of 5 nm. Therefore, one evident question to assess the practical utility of these designs would be to evaluate how random variations in the thickness of individual steps affect the overall performance of the HLIS. In order to answer this question, we did some calculations to investigate how errors in deposited thicknesses change the short circuit current of the multijunction solar cell. In these computational experiments we started from optimum HLIS using $Al_2O_3/TiO_2$ as optical materials and 5nm steps on 4J IMM solar cells for ambient glass and both spectra used in previous sections. We considered two scenarios, namely, a thickness variation suffering from a random Gaussian error with standard deviation of 20% –as reported in (Dong et al., 2018)– and a more favorable situation with a 10% standard deviation. The results of such calculations are summarized in **Table 1**. The approach was as follows: 1) we start with the optimum HLIS with steps of 5nm; 2) we create a "perturbed" HLIS by adding random Gaussian noise to the thickness of each step in the layer. The standard deviation of such Gaussian noise is either 10% or 20%, as indicated in the fourth column of Table 1; 3) we repeat step 2) ten times, thereby generating ten different "perturbed" designs for each case. In **Table 2**, as an example, we present the ten "perturbed" (with σ=20%) HLIS designs for 4JMM solar cells for AM1.5D and ambient glass using $Al_2O_3/TiO_2$ materials. The perturbation consists in the modification of the optimum thickness values for each later with the addition random of Gaussian noise with a standard deviation of 20%. The number of layers (from the interface with ambient) and material of HLIS design are indicated in first two columns. The column T0 is the target thicknesses in nanometers whereas T1-T10 are the resulting "perturbed" designs; 4) we calculate the photocurrent produced by each design and then the mean and variations compared to original $J_{SC}$ of the set (columns 5 and 6 in Table 1).

In short, the two rightmost columns in Table 1 give us an idea of how a HLIS design would perform when affected by random deviations of 10% or 20% from the optimum in the individual steps constituting the design. The results of Table 1 reflect that HLIS designs are surprisingly robust against moderate random variations. When the deviation in thickness is of 10% the loss in the resulting mean $J_{SC}$ is less than 0.26%, whereas when the deviation in thickness is of 20% the drop in the resulting mean $J_{SC}$ is less than 0.60% in the worst case. In addition, as shown in *Fig 5b*, the $J_{SC}$ of a DLAR design for 4JIMM with spectra AM1.5D and ambient glass is 12.64 mA/cm$^2$ while the $J_{SC}$ of HLIS design after 20% error in thickness is 12.95 mA/cm$^2$ and thus a 2.5% improvement is still obtained even after the addition of a 20% error in thickness compared to DLAR design. Analogously, the $J_{SC}$ of HLIS design even after the effect of a 20% error in thickness is higher 2.3% than that of a DLAR design (as shown in Fig 6b, 15.90 mA/cm$^2$) for spectra AM0. In summary, we can see from Table 1 that the impact of deposition tolerances is very small when the deposition uncertainty is 10% and even when it raises to 20%, the impact for $J_{SC}$ is also acceptable.

**Table 1.** Impact of ARC deposition tolerances of optimum HLIS design with 5nm slices using $Al_2O_3/TiO_2$ materials on short circuit currents density of 4J IMM solar cells.

| Spectrum | $J_{SC}$ [mA/cm$^2$] | Number of simulations | Std. Dev. in Thickness | Mean $J_{SC}$ [mA/cm$^2$] | $J_{SC}$ Variation |
|---|---|---|---|---|---|
| AM1.5D, Glass | 12.97 | 10 | 10% | 12.96 | -0.12% |
| | 12.97 | 10 | 20% | 12.95 | -0.17% |
| AM0, Glass | 16.36 | 10 | 10% | 16.32 | -0.26% |
| | 16.36 | 10 | 20% | 16.27 | -0.60% |

**Table 2.** Ten different "perturbed" HLIS designs obtained by adding random Gaussian noise with 20% standard deviation for 4J IMM solar cells for AM1.5D and glass using $Al_2O_3/TiO_2$.

| Layer | Material | T0 | T1 | T2 | T3 | T4 | T5 | T6 | T7 | T8 | T9 | T10 |
|---|---|---|---|---|---|---|---|---|---|---|---|---|
| 1 | TiO$_2$ | 5.00 | 4.35 | 5.86 | 4.53 | 2.50 | 4.40 | 4.18 | 5.32 | 6.13 | 4.94 | 5.81 |
| 2 | Al$_2$O$_3$ | 80.0 | 99.0 | 68.9 | 82.1 | 87.0 | 81.6 | 78.4 | 71.0 | 75.3 | 89.7 | 88.73 |
| 3 | TiO$_2$ | 5.00 | 3.39 | 5.45 | 4.71 | 3.60 | 5.56 | 5.34 | 4.69 | 6.26 | 5.11 | 3.95 |
| 4 | Al$_2$O$_3$ | 25.0 | 24.8 | 25.5 | 26.5 | 23.7 | 25.5 | 20.4 | 22.1 | 27.3 | 34.0 | 26.99 |
| 5 | TiO$_2$ | 5.00 | 3.05 | 5.83 | 5.40 | 5.16 | 4.10 | 4.71 | 3.97 | 6.17 | 5.31 | 4.25 |
| 6 | Al$_2$O$_3$ | 10.0 | 12.0 | 11.0 | 8.14 | 11.5 | 9.06 | 10.7 | 8.18 | 10.2 | 13.6 | 13.03 |
| 7 | TiO$_2$ | 5.00 | 5.86 | 5.90 | 4.82 | 4.73 | 4.88 | 3.16 | 4.79 | 4.34 | 4.28 | 4.97 |
| 8 | Al$_2$O$_3$ | 5.00 | 5.00 | 4.87 | 2.87 | 6.58 | 6.48 | 6.04 | 3.30 | 3.52 | 5.53 | 6.64 |
| 9 | TiO$_2$ | 5.00 | 4.93 | 4.85 | 6.15 | 4.52 | 4.14 | 7.42 | 5.61 | 5.16 | 4.74 | 4.57 |
| 10 | Al$_2$O$_3$ | 10.0 | 5.03 | 12.0 | 8.74 | 10.6 | 11.5 | 11.9 | 9.76 | 11.6 | 11.2 | 11.18 |
| 11 | TiO$_2$ | 5.00 | 5.58 | 2.88 | 3.80 | 5.66 | 5.31 | 4.68 | 5.70 | 4.71 | 5.59 | 4.94 |
| 12 | Al$_2$O$_3$ | 5.00 | 2.81 | 4.50 | 4.75 | 5.09 | 4.77 | 5.43 | 5.27 | 4.46 | 2.81 | 2.98 |
| 13 | TiO$_2$ | 15.0 | 8.04 | 11.1 | 10.7 | 17.6 | 11.8 | 11.8 | 16.4 | 14.0 | 11.0 | 12.05 |
| 14 | Al$_2$O$_3$ | 5.00 | 5.08 | 4.62 | 4.98 | 5.32 | 4.72 | 6.88 | 3.52 | 3.90 | 3.56 | 5.61 |
| 15 | TiO$_2$ | 25.0 | 20.2 | 28.2 | 22.2 | 21.0 | 24.5 | 29.7 | 19.9 | 22.5 | 27.0 | 24.73 |
| 16 | Al$_2$O$_3$ | 5.00 | 5.41 | 5.83 | 7.18 | 3.19 | 3.53 | 5.79 | 4.55 | 4.82 | 6.47 | 3.88 |
| 17 | TiO$_2$ | 5.00 | 17.0 | 11.9 | 18.4 | 20.5 | 15.5 | 12.3 | 15.3 | 15.1 | 14.0 | 13.12 |

**4.2. Effect of Obliquity of Incident Light on ARC Performance**

In our calculation, we found the HLIS designs outperform DLAR for normal incidence. However, in practice, the incident light is not always perpendicular to the surface of the antireflection coating and the performance of ARC at oblique incidence is important. (Espinet-González et al., 2012). Consequently, the HLIS performance at oblique incidence should also be considered. We investigated by simulations the effect of angle of incidence for HLIS and DLAR designs. We calculated the performance of a HLIS with 5nm slices and a DLAR at the angles of incidence from 0° to 90° with ambient glass for both spectra AM0 and AM1.5D using as optical materials $Al_2O_3/TiO_2$. In this calculation, we again used the transfer matrix method and the total transfer matrix of multilayers was calculated as the product of the transfer matrix of each layer in the order of light propagation (Centurioni, 2005). The reflectance is the average of the s-polarization and p-polarization components since for oblique incidence, the values of s-polarization and p-polarization are different (Sharma, 2019). The results are shown in *Fig 7*, where the blue line is for HLIS design while the red line is for DLAR design. *Fig 7*(a) and (b) show the calculation for triple junction solar cells, whist (c) and (d) are for quadruple junction solar cells. As shown, for HLIS and DLAR, the changes in short circuit current density are very small as the angle increases from 0° to 60° for both triple-junction solar cells and quadruple-junction solar cells. Therefore, the HLIS and DLAR show similar behavior for the oblique incident light. Comparing these two designs, the performance of HLIS is slightly better than the DLAR almost in the entire range of angle (from 0° to 85°). Briefly, the HLIS design has excellent performance at oblique incidence which provides more possibilities for practical applications.

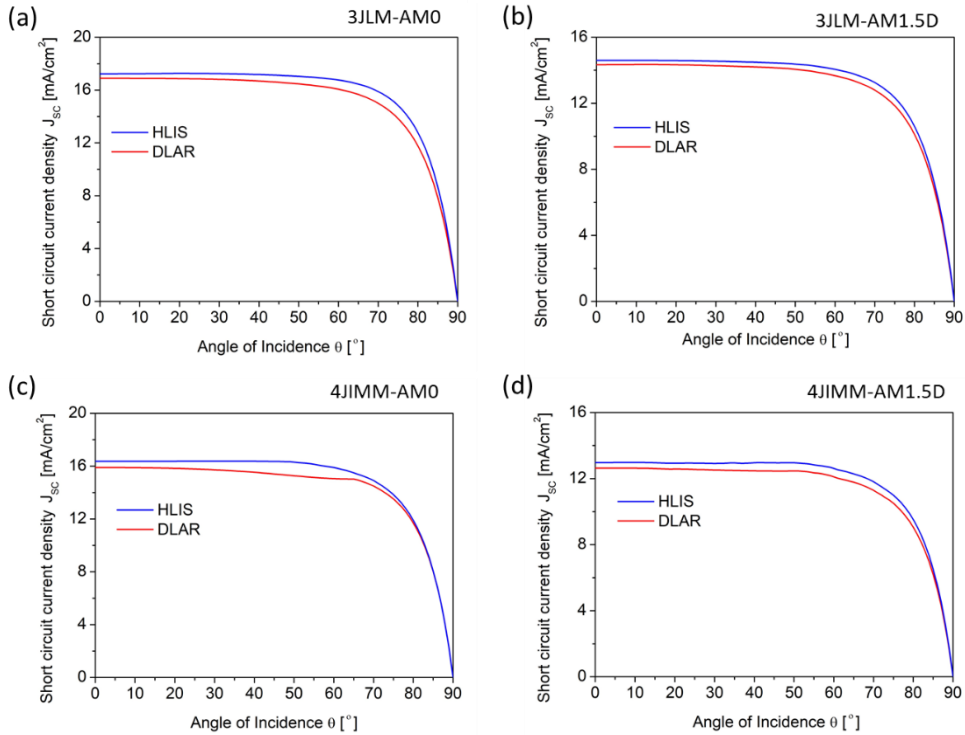

**Fig 7**. The performance of HLIS and DLAR designs at an oblique angle of incidence on triple-junction solar cells and quadruple-junction solar cells with ambient glass for both spectra AM0 and AM1.5D using $Al_2O_3/TiO_2$ as optical materials.

### 4.3. A Word on Deposition Techniques and Tools

The standard approach followed in many labs and industries to deposit antireflection coatings on III-V multijunction solar cells is Physical Vapor Deposition (PVD) (Mahan, 2000). In most cases e-beam evaporation is used (for $Al_2O_3/TiO_2$) or in some other cases thermal evaporation is an option ($MgF_2$/ZnS). In the typical configuration of these tools only one material can be evaporated at a time since there is only one active crucible or boat radiating material into the vacuum chamber. Such tool configuration works nicely for DLAR but would be unpractical to deposit a HLIS, which could need tens of steps of 5nm and frequent flipping from high to low index materials or vice versa (i.e. with continuous heating up/cooling down cycles to switch materials). However, this problem can be sorted out easily using a tool equipped with (at least) two radiating crucibles or boats blocked or activated by the corresponding shutters. The fast flipping between materials is then reduced to a problem of fast actuation over a mechanical shutter. Optical coaters implementing similar strategies are of common use in other industries (for coatings on glass for example) and even much more complex architectures have been proposed in the literature (Sullivan and Dobrowolski, 1993; Sullivan et al. 2000). This deposition technique of HLIS design will not increase costs since it neither requires sophisticated equipment nor increase the amount of deposition materials. As well as, the advantage of HLIS ARC is that only two materials are used which reduce the demand of new materials compared to a multi-layer step down ARC (e.g. triple or quadruple layer ARC). This simple technical solution together with the fact that HLIS seem to be largely immune to small variations in step thickness show promise for their practical applicability.

## 5. Conclusion

In this paper, we develop the so called high-low refractive index stacks design for antireflection coatings for multijunction solar cells. We have calculated the performance of HLIS with the same thickness and materials as optimum DLAR. We have presented the simulation results that the short circuit current density can be increased for triple-junction solar cells and quadruple-junction solar cells for AM1.5D and AM0 solar spectra when applying high low refractive index stacks antireflection coatings instead of optimum double layer antireflection coatings. For GaInP/Ga(In)As/Ge triple-junction solar cells, we got maximum improvement 1.82% with spectra AM1.5D and 2.16% with spectra AM0 comparing to DLAR. For quadruple-junction IMM GaInP/GaAs/GaInAs/GaInAs solar cells, we got considerable improvement, 4.64% with ambient air by using optical materials $Al_2O_3$/$TiO_2$ for AM1.5D and 4.52% with ambient air by using $MgF_2$/ZnS for AM0. The analysis of thickness uncertainty shows that when the deposition thickness uncertainty is <20%, the effect on the $J_{SC}$ of quadruple-junction cells is below 0.60%. A technical solution for depositing HILS has been proposed. Therefore, the potential of high low refractive index stacks designs is demonstrated.

**Acknowledgements**

**Funding**: This work was supported by the Spanish Ministerio de Economía y Competitividad through projects VIGNEMALE (RTI2018-094291-B-I00) and by the Comunidad de Madrid through the project ELOISE. I. García acknowledges the financial support from the Spanish Programa Estatal de Promoción del Talento y su Empleabilidad through a Ramón y Cajal [grant number RYC-2014-15621]. GuoJiao Hou acknowledges the financial support from the China Scholarship Council through [grant number 201804910474].